\documentclass[pra,reprint,amssymb,superscriptaddress]{revtex4-2}  
\usepackage{natbib}
\usepackage{times}
\usepackage{graphicx}  
\usepackage{graphics}
\usepackage{dcolumn}   
\usepackage{bm}        
\usepackage{amssymb}   
\usepackage{amsmath}
\usepackage{amssymb}
\usepackage{float}
\usepackage[colorlinks,linkcolor=blue,citecolor=blue]{hyperref}
\usepackage{epstopdf}
\usepackage{bookmark}
\usepackage{xcolor}
\usepackage{lipsum}
\newcommand\bra[2][]{#1\langle {#2} #1\rvert}
\newcommand\ket[2][]{#1\lvert {#2} #1\rangle}

\begin{document}

\widetext

\title{\textbf{Heralded Three-Photon Entanglement from a Single-Photon Source on a Photonic Chip}}

\author{Si Chen}
\author{Li-Chao Peng}
\author{Yong-Peng Guo}
\author{Xue-Mei Gu}
\author{Xing Ding}
\author{Run-Ze Liu}

\affiliation{%
    Hefei National Research Center for Physical Sciences at the Microscale and School of Physical Sciences, University of Science and Technology of China, Hefei 230026, China%
}
\affiliation{%
    Shanghai Research Center for Quantum Science and CAS Center for Excellence in Quantum Information and Quantum Physics, University of Science and Technology of China, Shanghai 201315, China%
}
\affiliation{%
    Hefei National Laboratory, University of Science and Technology of China, Hefei 230088, China%
}

\author{Xiang You}

\affiliation{%
    Hefei National Research Center for Physical Sciences at the Microscale and School of Physical Sciences, University of Science and Technology of China, Hefei 230026, China%
}
\affiliation{%
    Shanghai Research Center for Quantum Science and CAS Center for Excellence in Quantum Information and Quantum Physics, University of Science and Technology of China, Shanghai 201315, China%
}
\affiliation{%
    Hefei National Laboratory, University of Science and Technology of China, Hefei 230088, China%
}
\affiliation{%
    University of Science and Technology of China, School of Cyberspace Security, Hefei, China%
}

\author{Jian Qin}
\author{Yun-Fei Wang}
\author{Yu-Ming He}
\affiliation{%
    Hefei National Research Center for Physical Sciences at the Microscale and School of Physical Sciences, University of Science and Technology of China, Hefei 230026, China%
}
\affiliation{%
    Shanghai Research Center for Quantum Science and CAS Center for Excellence in Quantum Information and Quantum Physics, University of Science and Technology of China, Shanghai 201315, China%
}

\affiliation{%
    Hefei National Laboratory, University of Science and Technology of China, Hefei 230088, China%
}
\author{Jelmer J. Renema}
\affiliation{%
Adaptive Quantum Optics Group, Mesa+ Institute for Nanotechnology, University of Twente,
P.O. Box 217, 7500 AE Enschede, Netherlands%
}
\author{Yong-Heng Huo}
\author{Hui Wang}
\author{Chao-Yang Lu}
\author{Jian-Wei Pan}
\affiliation{%
    Hefei National Research Center for Physical Sciences at the Microscale and School of Physical Sciences, University of Science and Technology of China, Hefei 230026, China%
}
\affiliation{%
    Shanghai Research Center for Quantum Science and CAS Center for Excellence in Quantum Information and Quantum Physics, University of Science and Technology of China, Shanghai 201315, China%
}

\affiliation{%
    Hefei National Laboratory, University of Science and Technology of China, Hefei 230088, China%
}
\date{\today}

\begin{abstract}
    In the quest to build general-purpose photonic quantum computers, fusion-based quantum computation has risen to prominence as a promising strategy. This model allows a ballistic construction of large cluster states which are universal for quantum computation, in a scalable and loss-tolerant way without feed-forward, by fusing many small $n$-photon entangled resource states. However, a key obstacle to this architecture lies in efficiently generating the required essential resource states on photonic chips. One such critical seed state that has not yet been achieved is the heralded three-photon Greenberger-Horne-Zeilinger (3-GHZ) state. Here, we address this elementary resource gap, by reporting the first experimental realization of a heralded dual-rail encoded 3-GHZ state. Our implementation employs a low-loss and fully programmable photonic chip that manipulates six indistinguishable single photons of wavelengths in the telecommunication regime. Conditional on the heralding detection, we obtain the desired 3-GHZ state with a fidelity $0.573\pm0.024$. Our work marks an important step for the future fault-tolerant photonic quantum computing, leading to the acceleration of building a large-scale optical quantum computer.
\end{abstract}

\maketitle

Photon is a favourable candidate for universal quantum computing \cite{knill2001scheme,kok2007linear,pan2012multiphoton}, allowing several advantages such as room-temperature operation, negligible decoherence, and easy integration into existing fiber-optic-based telecommunications systems. Especially, the rapidly developed integrated optics makes it an appealing physical platform for large-scale fault-tolerant quantum computing \cite{o2007optical,ladd2010quantum,rudolph2017optimistic}.


Measurement-based quantum computing (MBQC) \cite{raussendorf2001one, raussendorf2003mbqc, walther2005experimental, briegel2009measurement}, where quantum algorithms are performed by making single qubit measurements on a large entangled state---usually called cluster states, holds significant potential for photonic systems based on linear optics. Photonic cluster states can be efficiently created from small resource states in a fusion mechanism \cite{browne2005resource,duan2005efficient}.
Later, a ballistic strategy for MBQC has been proposed \cite{kieling2007percolation, gimeno2015three,pant2019percolation}, enabling scalable and loss-tolerant generation of large cluster states by fusing many small entangled resource states without any feed-forward, which has subsequently renormalized to fusion-based quantum computation \cite{bartolucci2023fusion}. So far, the heralded three-photon Greenberger-Horne-Zeilinger state has been identified as the minimal initial entangled state \cite{bartolucci2023fusion}, serving as an essential building block for constructing large entangled cluster states \cite{gimeno2015three, zaidi2015near, rudolph2017optimistic}.

Deterministic generations of multiphoton cluster states from single-photon sources have been proposed \cite{lindner2009proposal,economou2010optically,gimeno2019deterministic} and implemented \cite{schwartz2016deterministic,lee2019quantum,thomas2022efficient,cogan2023deterministic}. However, high-quality deterministic preparation still remains challenging with existing technology. Alternatively, without detecting or destroying the photons, one can near-deterministically generate such entangled clusters with heralded 3-GHZ states, which can be obtained using six single photons \cite{varnava2008good,gubarev2020improved}.

Here, we report the first experimental demonstration of a heralded dual-encoded 3-GHZ state using six single photons manipulated in a photonic chip. A high-quality single-photon source based on a semiconductor quantum dot \cite{senellart2017high} embedded in an open microcavity is used to deterministically produce single photons that are converted to the telecommunication band with a quantum frequency converter \cite{weber2019two,you2022quantum}. These single photons are deterministically demultiplexed into six indistinguishable single-photon sources \cite{wang2017high,wang2019boson}, which are manipulated in a fully programmable photonic chip \cite{taballione2021universal}. Heraled by the detection of four output spatial modes with high-efficiency single-photon detectors, we obtain a heralded 3-GHZ state with a fidelity of $0.573\pm0.024$. Our work is an important step towards fault-tolerant scalable photonic quantum computation.

\begin{figure}%
    \centering
    \includegraphics[width=0.48\textwidth]{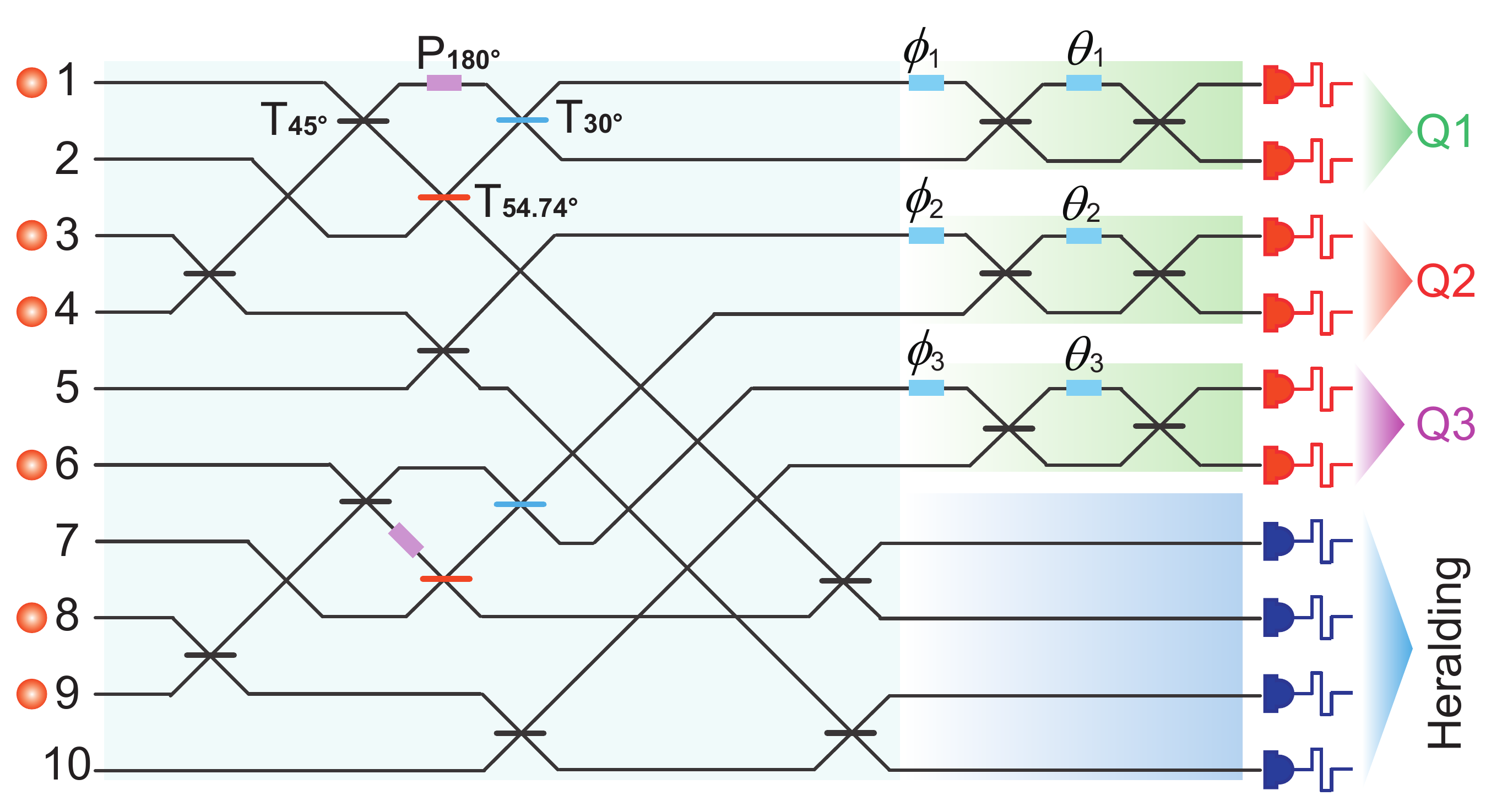}
    \caption{The heralded generation scheme for a 3-GHZ state. Six single photons are prepared at the input ports of a linear optical circuit, where we can describe the initial input quantum state as $\ket{\psi_{in}}=\ket{1011010110}$ with 0 and 1 being the number of photons in each mode. The photonic circuit (depicted in the light cyan background color), which contains twelve beams splitters of three different transmissions 1/2 (black), 1/4 (blue), 2/3 (orange) and two $\pi$ phase shifters (purple), implements a required unitary transformation $U$ for a 3-GHZ state. The 3-GHZ state in dual-rail encodings (a single photon appearing in only one of the two spatial modes describes a qubit such as $Q_{1}$, $Q_{2}$ and $Q_{3}$) is heralded in the modes (1-6) by a measurement pattern in the last four modes (7-10). Three phase-tunable Mach-Zehnder interferometers of phases $\varphi_i$ and $\theta_i$ ($i = 1, 2, 3$) are used to perform quantum state measurements on the created GHZ state.}\label{Fig:1}
\end{figure}

Fig.~\ref{Fig:1} illustrates a heralded generation of a dual-rail-encoded 3-GHZ state out of six single photons \cite{gubarev2020improved}. The scheme exploits a ten-mode linear optical circuit, which consists of twelve optical beams plitters with three distinct transmission coefficients and a pair of $\pi$-phase shifters. Six photons are injected into six specific input modes of the photonic circuit, preparing a number-basis initial state represented as $\ket{\psi_{in}}=a_1^{\dagger}a_3^{\dagger}a_4^{\dagger}a_6^{\dagger}a_8^{\dagger}a_9^{\dagger}\ket{0}^{\otimes 10}=\ket{1011010110}$.
Here, $a_i^{\dagger}$ refers to the creation operator in mode $i$, $\ket{0}$ symbolizes the vacuum state, and the numbers $0$ and $1$ indicate the number of photons occupied in each mode. The underlying unitary transformation $U$ of the circuit is optimally parameterized such that a particular measurement patterns in four ancillary modes (7-10) herald a desired 3-GHZ state at six output modes (1-6) \cite{gubarev2020improved}. Qubits $Q_{1}$, $Q_{2}$ and $Q_{3}$ are identified with output mode pairs (1,2), (3,4) and (5,6). This association uses a dual-rail encoding method, which signifies the presence of one single photon in either of the two spatial modes within each mode pair, e.g., $\ket{0}_{d}=\ket{10}$ and $\ket{1}_{d}=\ket{01}$. The GHZ state is heralded in modes (1-6) only when single photons are detected in both (7,8) and in just one of the (9,10) ports, with the other one remaining in a vacuum state. Each event has a success probability of $1/108$, resulting in an overall success rate of $1/54$ \cite{gubarev2020improved}. Using a phase-tunable Mach-Zehnder interferometers (MZIs) at each output mode pair, one can perform arbitrary local projective measurements for estimating the full state. Details of the underlying state evolution and state measurements are provided in Supplementary.

To prepare six single photons, we firstly use the state-of-the-art self-assembled InAs/GaAs quantum dot (QD), which is coupled to an polarized and tunable microcavity \cite{wang2019towards,tomm2021bright} and cooled down to $\sim$4 K. Under resonant pumping by a $\pi$-pulse laser with a repetition rate of $\sim$76 MHz, the QD emits $\sim$50 MHz polarized resonance fluorescence single photons at the end of the single-mode fiber. We measured second-order correlation of the photon source with a Hanbury Brown-Twiss (HBT) interferometer \cite{hanbury1956question}, and obtained $g^2(0)=0.028(8)$ at zero delay, which indicates a high single-photon purity of 97.2(8)\%. The single photon indistinguishability is tested using a Hong-Ou-Mandel (HOM) interferometer \cite{hong1987measurement}, yielding a visibility of 89(1)\% between two photons separated by $\sim$13 ns.

\begin{figure*}
    \centering
    \includegraphics[width=1.\textwidth]{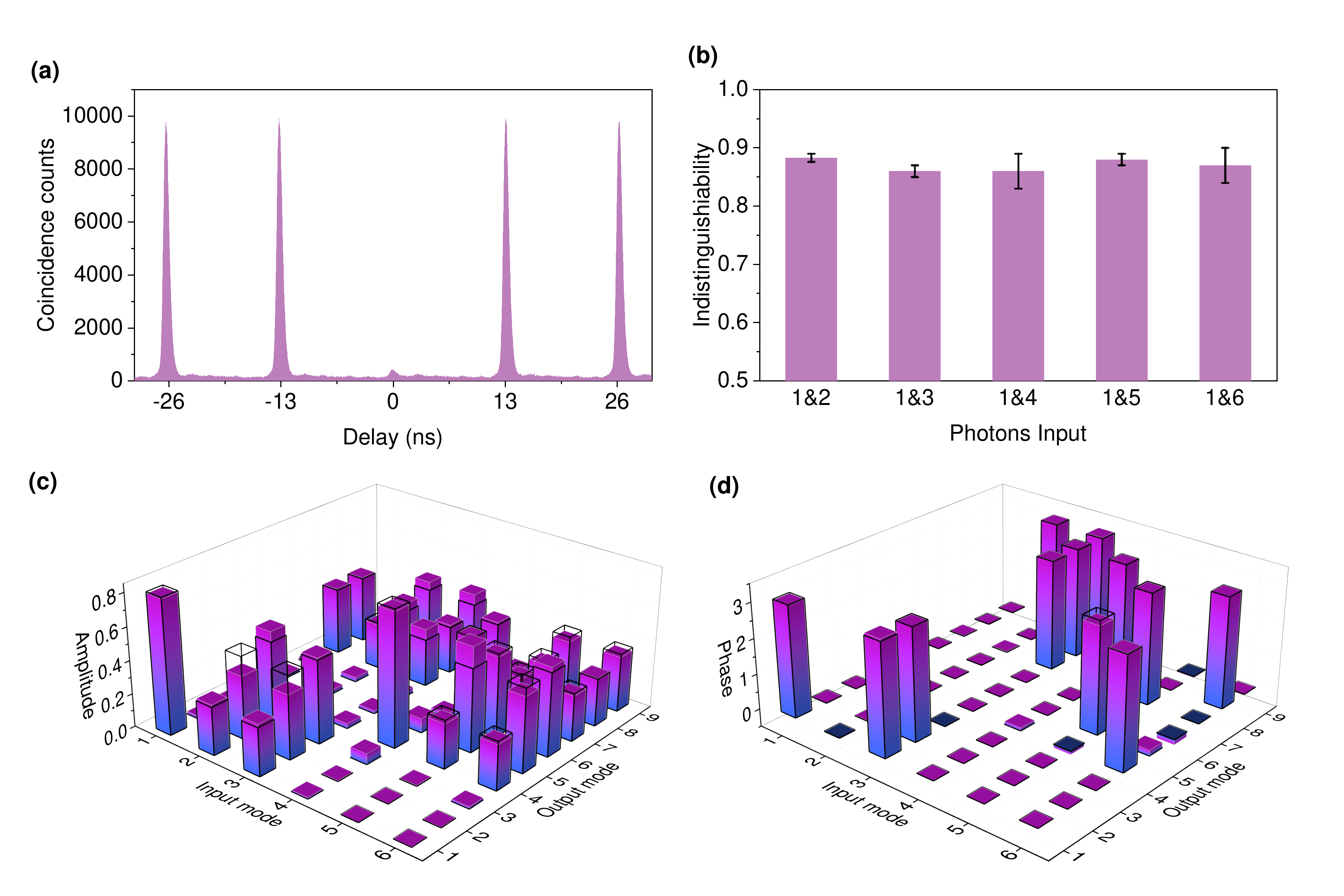}
    \caption{Characterization of the single-photon source and the photonic quantum circuit. (a), The purity of the single-photon source of telecom-wavelength is evaluated by a second-order correlation function $g^2(0)$. The observed anti-bunched peak at zero-time delay reveals $g^2(0)=0.026(6)$. (b), Measurement of the photon indistinguishability by performing a Hong-Ou-Mandel interferometer between photon 1 and photon 2 (3, 4, 5 or 6). The corresponding measured indistinguishability are $0.883(7)$, $0.86(1)$, $0.86(3)$, $0.88(1)$, $0.87(3)$, respectively. The error bars indicates one standard deviation, deduced from propagated Poissonian counting statistics of the raw photon detection events. (c)-(d), Amplitude (c) and Phase (d) of the ideal unitary transformation matrix (black wire frames) and their experimentally reconstructions (colored bars). We send six single photons into the used six input ports of the photonic circuit, and measure the normalized amplitude and phase for each input–output combination.}
    \label{Fig:3}
\end{figure*}

We then employ a quantum frequency converter (QFC) to transfer the near-infrared wavelength of the produced single photons to the preferable telecommunication regime \cite{you2022quantum}. For this purpose, we fabricate a periodically poled lithium niobate (PPLN) waveguide for difference-frequency generation process that can be adjusted by the wavelengths of the pump lasers. A continuous wave (CW) pump laser at $\sim$2060 nm and the QD-emitted single photons at $\sim$884.5 nm are then coupled into the PPLN waveguide, in which the difference frequency generation occurs, thus generating the output single photons at 1550 nm. By optimizing the waveguide coupling, and transmission and detection rate, we eventually achieve a overall single-photon conversion efficiency of $\sim$50\%. To test whether the converted photons still preserve the single-photon nature and their indistinguishability, we preform the HBT and HOM measurements on the photons after conversion. The purity of the single photons at 1550 nm stays $97.4(6)\%$, and the indistinguishability between photon 1 and photons 2, 3, 4, 5, 6 are respectively 0.883(7), 0.86(1), 0.86(3), 0.88(1), 0.87(3), as shown in Fig.~\ref{Fig:3} (a) and (b).

\begin{figure*}
    \centering
    \includegraphics[width=1\textwidth]{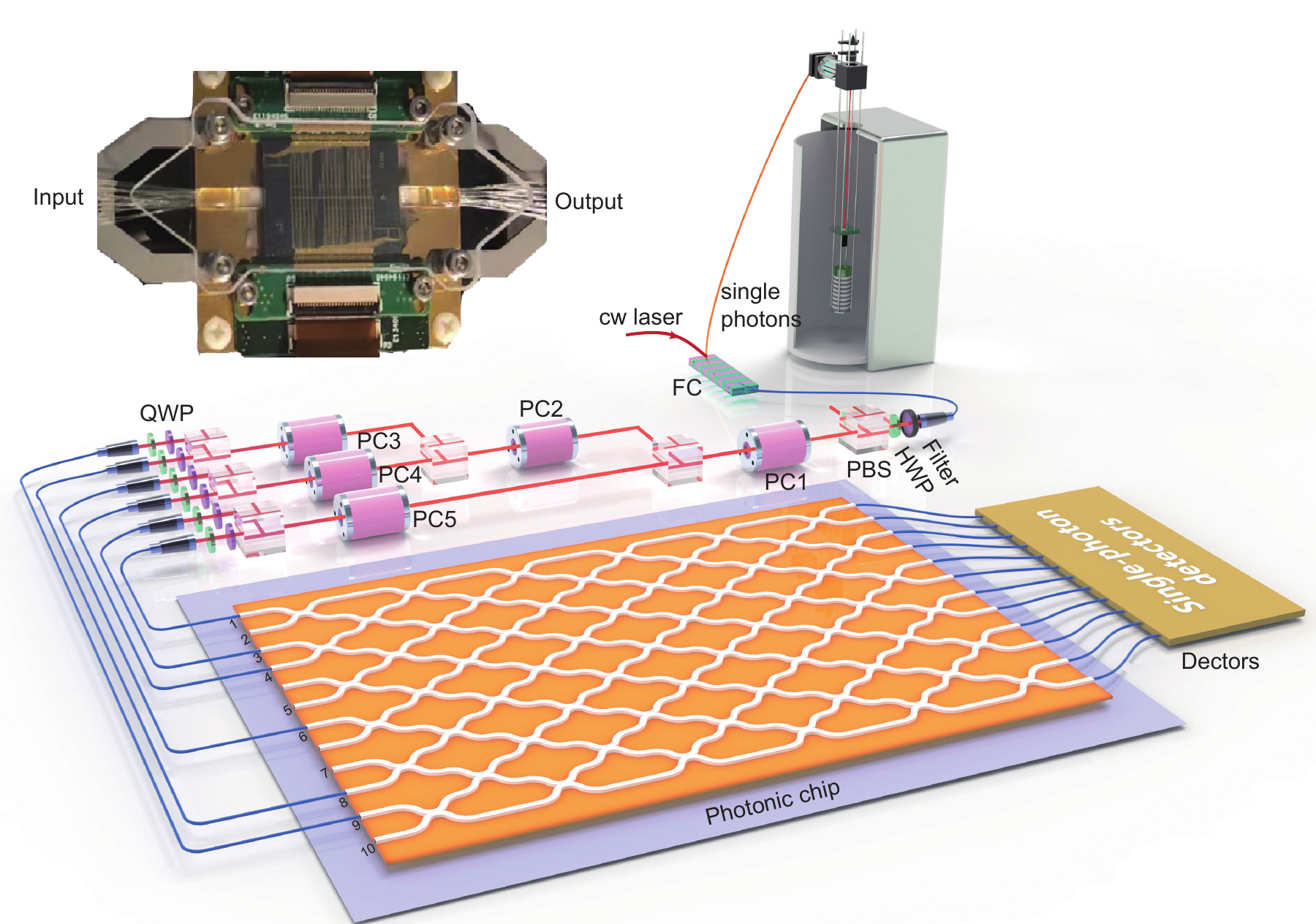}
    \caption{The experimental setup. A single InAs/GaAs quantum dot, resonantly coupled to an open microcavity, is used to produce pulsed resonance fluorescence single photons. The wavelength of these single photons in near-infrared is then converted into telecom-wavelength of 1550 nm with a quantum frequency converter by coupling a continuous wave laser and single photons into a periodically poled lithium niobate waveguide. Five pairs of Pockels cells and polarizing beam splitters are employed to actively translate a stream of single photons into six spatial modes. Optical fibers of different lengths are used to precisely compensate time delays. The six indistinguishable single photons are then fed into a 12-modes programmable linear-optical quantum circuit, which allows the preparation and measurement of a heralded three-qubit dual-rail encoded GHZ state. The single photons at the output ports of the photonic circuit is detected by superconducting nanowire single-photon detectors. All coincidence are recorded by a coincidence count unit (not shown).}
    \label{Fig:2}
\end{figure*}

The converted single-photon stream is then deterministically demultiplexed into six spatial modes using a tree-like structure demultiplexer constructed by five pairs of Pockels cells (PCs) and polarizing beam splitters (PBSs). The PCs, synchronized to the laser pulses and operated at a repetition rate of $\sim$705 kHz, actively control the photon polarization when loaded with high-voltage electrical pulses. The measured average optical switches efficiency is $\sim$77\%, which is mainly due to the coupling efficiency and propagation loss. With the help of six single-mode fibers of different lengths and translation stages, we precisely compensate the relative time delays of the six single photons such that they can simultaneously arrive at the input ports of the photonic circuit.

To realize the functional design of the unitary transformation in Fig.~\ref{Fig:1}, we employ a photonic chip that is low-loss and fully programmable \cite{taballione2021universal}. The circuit is based on stoichiometric silicon nitride waveguides which are fabricated for single-mode operation at a wavelength of 1550 nm. It consists 12 input and output spatial modes that are interconnected through an arrangement of adjustable beam splitters and thermo-optic phase shifters, as shown in Fig.\ref{Fig:2} (details about the photonic circuit, please refer to Ref.~\cite{taballione2021universal}). To achieve a heralded 3-GHZ state, six single photons are injected into six inputs (1, 3, 4, 6, 8, 9) of the circuit and propagate through the circuit. The heralded outputs are 7, 8, 9 and 10. At each heralded output (7, 8, 9), two superconducting nanowire single-photon detectors (SNSPDs) are employed and act as a pseudo-photon-number-detector that can resolve up to two photons. When each mode (7, 8, 9) contains a single photon and mode 10 has vacuum state, one can obtain a heralded GHZ generation for three dual-rail encoded qubits defined in the modes $Q1=(1,2)$, $Q2=(3,4)$ and $Q3=(5,6)$.

We then send the six photons one by one and measure the output distribution at the nine output modes (1-9) for analyzing the quality of the photonic circuit. For each input-output combination, we implement a Mach-Zehnder-type coherence measurement to record the corresponding phase. The normalized amplitude and measured phase compared to their theoretical distributions are summarized in Fig.~\ref{Fig:3} (c) and (d).

\begin{figure}
    \centering
    \includegraphics[width=0.45\textwidth]{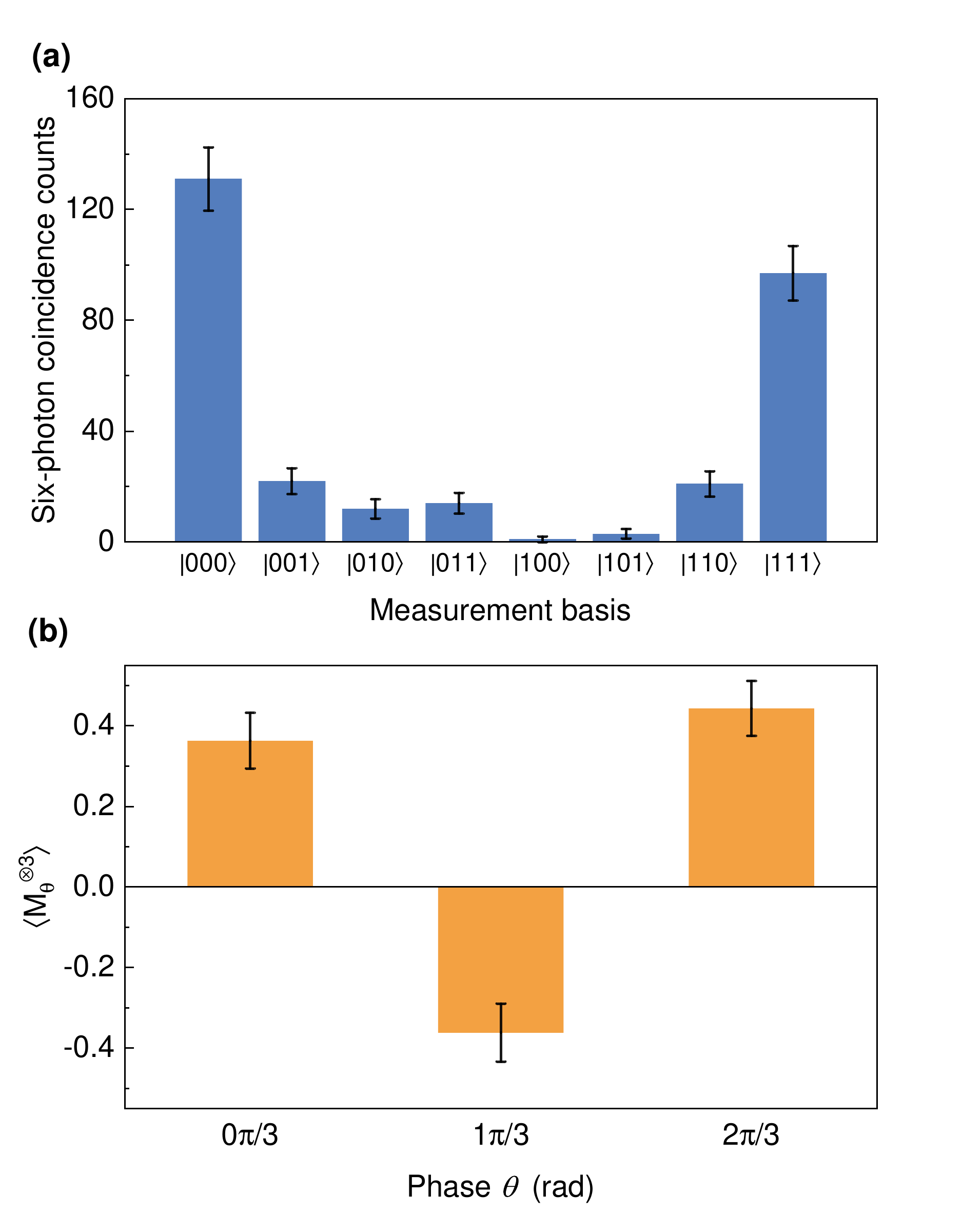}
    \caption{Experimental results of the generated heralded three-qubit GHZ state. (a), Six-photon coincidence counts measured in the dual-rail basis $\ket{0}_{d}/\ket{1}_{d}$ accumulated for 23 h. (b), Expectation values of $\langle M^{\otimes 3}_{\theta}\rangle$ with phase shift being $\theta=k\pi/3$ ($k=0,1,2$), each obtained from the measurement in the basis of $(\ket{0}_{d}\pm e^{i\theta}\ket{1}_{d})/\sqrt{2}$. Error bars indicate one standard deviation calculated from Poissonian counting statistics of the raw detection events.}
    \label{Fig:4}
\end{figure}

To analyze the generated heralded three-photon GHZ state in dual-rail encoding, we use the phase-tunable MZIs to perform any local projective measurements on the single photons. The transformation matrixes of these local measurements are compiled in the whole circuit. We then collect the six-photon coincidence counts at the used ten outputs in which each output mode (7, 8, 9) contains only one photon. To validate the three-qubit GHZ entanglement, we first measure the six-photon events in the $\ket{0}_{d}/\ket{1}_{d}$ basis (see data in Fig.~\ref{Fig:4} (a)) to calculate the population of $(\ket{0}_{d}\bra{0}_{d})^{\otimes3}+(\ket{1}_{d}\bra{1}_{d})^{\otimes3}$ over all the possible $2^{3}$ combinations, leading to a population $P=0.758 \pm 0.025$. We further estimate the expectation value of the observable $M^{\bigotimes N}_{\theta}=(\cos \theta\hat{\sigma}_x+\sin\theta\hat{\sigma}_y)^{\bigotimes N}$, where $\theta=k\pi/3$ ($k=0,1,2$) and $\hat{\sigma}_x$,$ \hat{\sigma}_y$ are Pauli matrices. The coherence of the three-qubit GHZ state is defined by the off-diagonal element of its density matrix and can be calculated by $C=(1/3)\sum_{k=0}^{2}(-1)^k\langle M_{k\pi/3}^{\bigotimes3}\rangle$, which is $C=0.389\pm0.040$ (see data in Fig.~\ref{Fig:4} (b)). The state fidelity can be directly estimated by $F=(P+C)/2=0.573\pm0.024$, which surpasses the classical threshold of 0.5 by more than 3 standard deviations and is sufficient to show the presence of entanglement \cite{guhne2009entanglement}.

In summary, we have demonstrated, for the first time, the heralded three-photon GHZ state using six photons from a high-quality quantum-dot single-photon source and a 12-mode fully programmable photonic chip. It should be noted that this experiment is unrealistic for commonly used spontaneous parametric down-conversion sources \cite{zhong201812}, since the expected count rate will be four orders of magnitude smaller than this experiment, showing a huge advantage of deterministic quantum-dot single-photon sources. Our demonstrated three-qubit GHZ state is heralded and has a heralding efficiency that could reach to one in principle, which is the main building block for fusion-based photonic quantum computing. Here, the heralding efficiency is defined as the probability of successfully achieving a desired GHZ state when there are coincidences of desired heralding detectors. In our experiment, we obtain a heralding efficiency of $\sim$0.0005 after the correction of detectors' imperfections (see Supplementary). This heralding efficiency can be significantly improved in the near future, by further promoting the efficiency of single-photon sources and photonic circuit to enable the large-scale photonic quantum computing.

~\\S. C. and L.-C P. contributed equally to this work.

\nocite{*}
\bibliographystyle{apsrev4-2}
\bibliography{ref}

\begin{thebibliography}{38}%
\makeatletter
\providecommand \@ifxundefined [1]{%
 \@ifx{#1\undefined}
}%
\providecommand \@ifnum [1]{%
 \ifnum #1\expandafter \@firstoftwo
 \else \expandafter \@secondoftwo
 \fi
}%
\providecommand \@ifx [1]{%
 \ifx #1\expandafter \@firstoftwo
 \else \expandafter \@secondoftwo
 \fi
}%
\providecommand \natexlab [1]{#1}%
\providecommand \enquote  [1]{``#1''}%
\providecommand \bibnamefont  [1]{#1}%
\providecommand \bibfnamefont [1]{#1}%
\providecommand \citenamefont [1]{#1}%
\providecommand \href@noop [0]{\@secondoftwo}%
\providecommand \href [0]{\begingroup \@sanitize@url \@href}%
\providecommand \@href[1]{\@@startlink{#1}\@@href}%
\providecommand \@@href[1]{\endgroup#1\@@endlink}%
\providecommand \@sanitize@url [0]{\catcode `\\12\catcode `\$12\catcode
  `\&12\catcode `\#12\catcode `\^12\catcode `\_12\catcode `\%12\relax}%
\providecommand \@@startlink[1]{}%
\providecommand \@@endlink[0]{}%
\providecommand \url  [0]{\begingroup\@sanitize@url \@url }%
\providecommand \@url [1]{\endgroup\@href {#1}{\urlprefix }}%
\providecommand \urlprefix  [0]{URL }%
\providecommand \Eprint [0]{\href }%
\providecommand \doibase [0]{https://doi.org/}%
\providecommand \selectlanguage [0]{\@gobble}%
\providecommand \bibinfo  [0]{\@secondoftwo}%
\providecommand \bibfield  [0]{\@secondoftwo}%
\providecommand \translation [1]{[#1]}%
\providecommand \BibitemOpen [0]{}%
\providecommand \bibitemStop [0]{}%
\providecommand \bibitemNoStop [0]{.\EOS\space}%
\providecommand \EOS [0]{\spacefactor3000\relax}%
\providecommand \BibitemShut  [1]{\csname bibitem#1\endcsname}%
\let\auto@bib@innerbib\@empty
\bibitem [{\citenamefont {Knill}\ \emph {et~al.}(2001)\citenamefont {Knill},
  \citenamefont {Laflamme},\ and\ \citenamefont {Milburn}}]{knill2001scheme}%
  \BibitemOpen
  \bibfield  {author} {\bibinfo {author} {\bibfnamefont {E.}~\bibnamefont
  {Knill}}, \bibinfo {author} {\bibfnamefont {R.}~\bibnamefont {Laflamme}},\
  and\ \bibinfo {author} {\bibfnamefont {G.~J.}\ \bibnamefont {Milburn}},\
  }\href {https://www.nature.com/articles/35051009} {\bibfield  {journal}
  {\bibinfo  {journal} {Nature}\ }\textbf {\bibinfo {volume} {409}},\ \bibinfo
  {pages} {46} (\bibinfo {year} {2001})}\BibitemShut {NoStop}%
\bibitem [{\citenamefont {Kok}\ \emph {et~al.}(2007)\citenamefont {Kok},
  \citenamefont {Munro}, \citenamefont {Nemoto}, \citenamefont {Ralph},
  \citenamefont {Dowling},\ and\ \citenamefont {Milburn}}]{kok2007linear}%
  \BibitemOpen
  \bibfield  {author} {\bibinfo {author} {\bibfnamefont {P.}~\bibnamefont
  {Kok}}, \bibinfo {author} {\bibfnamefont {W.~J.}\ \bibnamefont {Munro}},
  \bibinfo {author} {\bibfnamefont {K.}~\bibnamefont {Nemoto}}, \bibinfo
  {author} {\bibfnamefont {T.~C.}\ \bibnamefont {Ralph}}, \bibinfo {author}
  {\bibfnamefont {J.~P.}\ \bibnamefont {Dowling}},\ and\ \bibinfo {author}
  {\bibfnamefont {G.~J.}\ \bibnamefont {Milburn}},\ }\href
  {https://journals.aps.org/rmp/abstract/10.1103/RevModPhys.79.135} {\bibfield
  {journal} {\bibinfo  {journal} {Reviews of Modern Physics}\ }\textbf
  {\bibinfo {volume} {79}},\ \bibinfo {pages} {135} (\bibinfo {year}
  {2007})}\BibitemShut {NoStop}%
\bibitem [{\citenamefont {Pan}\ \emph {et~al.}(2012)\citenamefont {Pan},
  \citenamefont {Chen}, \citenamefont {Lu}, \citenamefont {Weinfurter},
  \citenamefont {Zeilinger},\ and\ \citenamefont
  {{\.Z}ukowski}}]{pan2012multiphoton}%
  \BibitemOpen
  \bibfield  {author} {\bibinfo {author} {\bibfnamefont {J.-W.}\ \bibnamefont
  {Pan}}, \bibinfo {author} {\bibfnamefont {Z.-B.}\ \bibnamefont {Chen}},
  \bibinfo {author} {\bibfnamefont {C.-Y.}\ \bibnamefont {Lu}}, \bibinfo
  {author} {\bibfnamefont {H.}~\bibnamefont {Weinfurter}}, \bibinfo {author}
  {\bibfnamefont {A.}~\bibnamefont {Zeilinger}},\ and\ \bibinfo {author}
  {\bibfnamefont {M.}~\bibnamefont {{\.Z}ukowski}},\ }\href
  {https://journals.aps.org/rmp/abstract/10.1103/RevModPhys.84.777} {\bibfield
  {journal} {\bibinfo  {journal} {Reviews of Modern Physics}\ }\textbf
  {\bibinfo {volume} {84}},\ \bibinfo {pages} {777} (\bibinfo {year}
  {2012})}\BibitemShut {NoStop}%
\bibitem [{\citenamefont {O'brien}(2007)}]{o2007optical}%
  \BibitemOpen
  \bibfield  {author} {\bibinfo {author} {\bibfnamefont {J.~L.}\ \bibnamefont
  {O'brien}},\ }\href {https://www.science.org/doi/abs/10.1126/science.1142892}
  {\bibfield  {journal} {\bibinfo  {journal} {Science}\ }\textbf {\bibinfo
  {volume} {318}},\ \bibinfo {pages} {1567} (\bibinfo {year}
  {2007})}\BibitemShut {NoStop}%
\bibitem [{\citenamefont {Ladd}\ \emph {et~al.}(2010)\citenamefont {Ladd},
  \citenamefont {Jelezko}, \citenamefont {Laflamme}, \citenamefont {Nakamura},
  \citenamefont {Monroe},\ and\ \citenamefont {O’Brien}}]{ladd2010quantum}%
  \BibitemOpen
  \bibfield  {author} {\bibinfo {author} {\bibfnamefont {T.~D.}\ \bibnamefont
  {Ladd}}, \bibinfo {author} {\bibfnamefont {F.}~\bibnamefont {Jelezko}},
  \bibinfo {author} {\bibfnamefont {R.}~\bibnamefont {Laflamme}}, \bibinfo
  {author} {\bibfnamefont {Y.}~\bibnamefont {Nakamura}}, \bibinfo {author}
  {\bibfnamefont {C.}~\bibnamefont {Monroe}},\ and\ \bibinfo {author}
  {\bibfnamefont {J.~L.}\ \bibnamefont {O’Brien}},\ }\href
  {https://www.nature.com/articles/nature08812} {\bibfield  {journal} {\bibinfo
   {journal} {Nature}\ }\textbf {\bibinfo {volume} {464}},\ \bibinfo {pages}
  {45} (\bibinfo {year} {2010})}\BibitemShut {NoStop}%
\bibitem [{\citenamefont {Rudolph}(2017)}]{rudolph2017optimistic}%
  \BibitemOpen
  \bibfield  {author} {\bibinfo {author} {\bibfnamefont {T.}~\bibnamefont
  {Rudolph}},\ }\href {https://doi.org/10.1063/1.4976737} {\bibfield  {journal}
  {\bibinfo  {journal} {APL Photonics}\ }\textbf {\bibinfo {volume} {2}},\
  \bibinfo {pages} {030901} (\bibinfo {year} {2017})}\BibitemShut {NoStop}%
\bibitem [{\citenamefont {Raussendorf}\ and\ \citenamefont
  {Briegel}(2001)}]{raussendorf2001one}%
  \BibitemOpen
  \bibfield  {author} {\bibinfo {author} {\bibfnamefont {R.}~\bibnamefont
  {Raussendorf}}\ and\ \bibinfo {author} {\bibfnamefont {H.~J.}\ \bibnamefont
  {Briegel}},\ }\href {https://doi.org/10.1103/PhysRevLett.86.5188} {\bibfield
  {journal} {\bibinfo  {journal} {Phys. Rev. Lett.}\ }\textbf {\bibinfo
  {volume} {86}},\ \bibinfo {pages} {5188} (\bibinfo {year}
  {2001})}\BibitemShut {NoStop}%
\bibitem [{\citenamefont {Raussendorf}\ \emph {et~al.}(2003)\citenamefont
  {Raussendorf}, \citenamefont {Browne},\ and\ \citenamefont
  {Briegel}}]{raussendorf2003mbqc}%
  \BibitemOpen
  \bibfield  {author} {\bibinfo {author} {\bibfnamefont {R.}~\bibnamefont
  {Raussendorf}}, \bibinfo {author} {\bibfnamefont {D.~E.}\ \bibnamefont
  {Browne}},\ and\ \bibinfo {author} {\bibfnamefont {H.~J.}\ \bibnamefont
  {Briegel}},\ }\href {https://doi.org/10.1103/PhysRevA.68.022312} {\bibfield
  {journal} {\bibinfo  {journal} {Phys. Rev. A}\ }\textbf {\bibinfo {volume}
  {68}},\ \bibinfo {pages} {022312} (\bibinfo {year} {2003})}\BibitemShut
  {NoStop}%
\bibitem [{\citenamefont {Walther}\ \emph {et~al.}(2005)\citenamefont
  {Walther}, \citenamefont {Resch}, \citenamefont {Rudolph}, \citenamefont
  {Schenck}, \citenamefont {Weinfurter}, \citenamefont {Vedral}, \citenamefont
  {Aspelmeyer},\ and\ \citenamefont {Zeilinger}}]{walther2005experimental}%
  \BibitemOpen
  \bibfield  {author} {\bibinfo {author} {\bibfnamefont {P.}~\bibnamefont
  {Walther}}, \bibinfo {author} {\bibfnamefont {K.~J.}\ \bibnamefont {Resch}},
  \bibinfo {author} {\bibfnamefont {T.}~\bibnamefont {Rudolph}}, \bibinfo
  {author} {\bibfnamefont {E.}~\bibnamefont {Schenck}}, \bibinfo {author}
  {\bibfnamefont {H.}~\bibnamefont {Weinfurter}}, \bibinfo {author}
  {\bibfnamefont {V.}~\bibnamefont {Vedral}}, \bibinfo {author} {\bibfnamefont
  {M.}~\bibnamefont {Aspelmeyer}},\ and\ \bibinfo {author} {\bibfnamefont
  {A.}~\bibnamefont {Zeilinger}},\ }\href
  {https://www.nature.com/articles/nature03347} {\bibfield  {journal} {\bibinfo
   {journal} {Nature}\ }\textbf {\bibinfo {volume} {434}},\ \bibinfo {pages}
  {169} (\bibinfo {year} {2005})}\BibitemShut {NoStop}%
\bibitem [{\citenamefont {Briegel}\ \emph {et~al.}(2009)\citenamefont
  {Briegel}, \citenamefont {Browne}, \citenamefont {D{\"u}r}, \citenamefont
  {Raussendorf},\ and\ \citenamefont {Van~den Nest}}]{briegel2009measurement}%
  \BibitemOpen
  \bibfield  {author} {\bibinfo {author} {\bibfnamefont {H.~J.}\ \bibnamefont
  {Briegel}}, \bibinfo {author} {\bibfnamefont {D.~E.}\ \bibnamefont {Browne}},
  \bibinfo {author} {\bibfnamefont {W.}~\bibnamefont {D{\"u}r}}, \bibinfo
  {author} {\bibfnamefont {R.}~\bibnamefont {Raussendorf}},\ and\ \bibinfo
  {author} {\bibfnamefont {M.}~\bibnamefont {Van~den Nest}},\ }\href
  {https://www.nature.com/articles/nphys1157} {\bibfield  {journal} {\bibinfo
  {journal} {Nature Physics}\ }\textbf {\bibinfo {volume} {5}},\ \bibinfo
  {pages} {19} (\bibinfo {year} {2009})}\BibitemShut {NoStop}%
\bibitem [{\citenamefont {Browne}\ and\ \citenamefont
  {Rudolph}(2005)}]{browne2005resource}%
  \BibitemOpen
  \bibfield  {author} {\bibinfo {author} {\bibfnamefont {D.~E.}\ \bibnamefont
  {Browne}}\ and\ \bibinfo {author} {\bibfnamefont {T.}~\bibnamefont
  {Rudolph}},\ }\href {https://doi.org/10.1103/PhysRevLett.95.010501}
  {\bibfield  {journal} {\bibinfo  {journal} {Phys. Rev. Lett.}\ }\textbf
  {\bibinfo {volume} {95}},\ \bibinfo {pages} {010501} (\bibinfo {year}
  {2005})}\BibitemShut {NoStop}%
\bibitem [{\citenamefont {Duan}\ and\ \citenamefont
  {Raussendorf}(2005)}]{duan2005efficient}%
  \BibitemOpen
  \bibfield  {author} {\bibinfo {author} {\bibfnamefont {L.-M.}\ \bibnamefont
  {Duan}}\ and\ \bibinfo {author} {\bibfnamefont {R.}~\bibnamefont
  {Raussendorf}},\ }\href
  {https://journals.aps.org/prl/abstract/10.1103/PhysRevLett.95.080503}
  {\bibfield  {journal} {\bibinfo  {journal} {Phys. Rev. Lett.}\ }\textbf
  {\bibinfo {volume} {95}},\ \bibinfo {pages} {080503} (\bibinfo {year}
  {2005})}\BibitemShut {NoStop}%
\bibitem [{\citenamefont {Kieling}\ \emph {et~al.}(2007)\citenamefont
  {Kieling}, \citenamefont {Rudolph},\ and\ \citenamefont
  {Eisert}}]{kieling2007percolation}%
  \BibitemOpen
  \bibfield  {author} {\bibinfo {author} {\bibfnamefont {K.}~\bibnamefont
  {Kieling}}, \bibinfo {author} {\bibfnamefont {T.}~\bibnamefont {Rudolph}},\
  and\ \bibinfo {author} {\bibfnamefont {J.}~\bibnamefont {Eisert}},\ }\href
  {https://doi.org/10.1103/PhysRevLett.99.130501} {\bibfield  {journal}
  {\bibinfo  {journal} {Phys. Rev. Lett.}\ }\textbf {\bibinfo {volume} {99}},\
  \bibinfo {pages} {130501} (\bibinfo {year} {2007})}\BibitemShut {NoStop}%
\bibitem [{\citenamefont {Gimeno-Segovia}\ \emph {et~al.}(2015)\citenamefont
  {Gimeno-Segovia}, \citenamefont {Shadbolt}, \citenamefont {Browne},\ and\
  \citenamefont {Rudolph}}]{gimeno2015three}%
  \BibitemOpen
  \bibfield  {author} {\bibinfo {author} {\bibfnamefont {M.}~\bibnamefont
  {Gimeno-Segovia}}, \bibinfo {author} {\bibfnamefont {P.}~\bibnamefont
  {Shadbolt}}, \bibinfo {author} {\bibfnamefont {D.~E.}\ \bibnamefont
  {Browne}},\ and\ \bibinfo {author} {\bibfnamefont {T.}~\bibnamefont
  {Rudolph}},\ }\href {https://doi.org/10.1103/PhysRevLett.115.020502}
  {\bibfield  {journal} {\bibinfo  {journal} {Phys. Rev. Lett.}\ }\textbf
  {\bibinfo {volume} {115}},\ \bibinfo {pages} {020502} (\bibinfo {year}
  {2015})}\BibitemShut {NoStop}%
\bibitem [{\citenamefont {Pant}\ \emph {et~al.}(2019)\citenamefont {Pant},
  \citenamefont {Towsley}, \citenamefont {Englund},\ and\ \citenamefont
  {Guha}}]{pant2019percolation}%
  \BibitemOpen
  \bibfield  {author} {\bibinfo {author} {\bibfnamefont {M.}~\bibnamefont
  {Pant}}, \bibinfo {author} {\bibfnamefont {D.}~\bibnamefont {Towsley}},
  \bibinfo {author} {\bibfnamefont {D.}~\bibnamefont {Englund}},\ and\ \bibinfo
  {author} {\bibfnamefont {S.}~\bibnamefont {Guha}},\ }\href
  {https://www.nature.com/articles/s41467-019-08948-x} {\bibfield  {journal}
  {\bibinfo  {journal} {Nature Communications}\ }\textbf {\bibinfo {volume}
  {10}},\ \bibinfo {pages} {1070} (\bibinfo {year} {2019})}\BibitemShut
  {NoStop}%
\bibitem [{\citenamefont {Bartolucci}\ \emph {et~al.}(2023)\citenamefont
  {Bartolucci}, \citenamefont {Birchall}, \citenamefont {Bombin}, \citenamefont
  {Cable}, \citenamefont {Dawson}, \citenamefont {Gimeno-Segovia},
  \citenamefont {Johnston}, \citenamefont {Kieling}, \citenamefont {Nickerson},
  \citenamefont {Pant} \emph {et~al.}}]{bartolucci2023fusion}%
  \BibitemOpen
  \bibfield  {author} {\bibinfo {author} {\bibfnamefont {S.}~\bibnamefont
  {Bartolucci}}, \bibinfo {author} {\bibfnamefont {P.}~\bibnamefont
  {Birchall}}, \bibinfo {author} {\bibfnamefont {H.}~\bibnamefont {Bombin}},
  \bibinfo {author} {\bibfnamefont {H.}~\bibnamefont {Cable}}, \bibinfo
  {author} {\bibfnamefont {C.}~\bibnamefont {Dawson}}, \bibinfo {author}
  {\bibfnamefont {M.}~\bibnamefont {Gimeno-Segovia}}, \bibinfo {author}
  {\bibfnamefont {E.}~\bibnamefont {Johnston}}, \bibinfo {author}
  {\bibfnamefont {K.}~\bibnamefont {Kieling}}, \bibinfo {author} {\bibfnamefont
  {N.}~\bibnamefont {Nickerson}}, \bibinfo {author} {\bibfnamefont
  {M.}~\bibnamefont {Pant}}, \emph {et~al.},\ }\href
  {https://doi.org/10.1038/s41467-023-36493-1} {\bibfield  {journal} {\bibinfo
  {journal} {Nature Communications}\ }\textbf {\bibinfo {volume} {14}},\
  \bibinfo {pages} {912} (\bibinfo {year} {2023})}\BibitemShut {NoStop}%
\bibitem [{\citenamefont {Zaidi}\ \emph {et~al.}(2015)\citenamefont {Zaidi},
  \citenamefont {Dawson}, \citenamefont {van Loock},\ and\ \citenamefont
  {Rudolph}}]{zaidi2015near}%
  \BibitemOpen
  \bibfield  {author} {\bibinfo {author} {\bibfnamefont {H.~A.}\ \bibnamefont
  {Zaidi}}, \bibinfo {author} {\bibfnamefont {C.}~\bibnamefont {Dawson}},
  \bibinfo {author} {\bibfnamefont {P.}~\bibnamefont {van Loock}},\ and\
  \bibinfo {author} {\bibfnamefont {T.}~\bibnamefont {Rudolph}},\ }\href
  {https://doi.org/10.1103/PhysRevA.91.042301} {\bibfield  {journal} {\bibinfo
  {journal} {Phys. Rev. A}\ }\textbf {\bibinfo {volume} {91}},\ \bibinfo
  {pages} {042301} (\bibinfo {year} {2015})}\BibitemShut {NoStop}%
\bibitem [{\citenamefont {Lindner}\ and\ \citenamefont
  {Rudolph}(2009)}]{lindner2009proposal}%
  \BibitemOpen
  \bibfield  {author} {\bibinfo {author} {\bibfnamefont {N.~H.}\ \bibnamefont
  {Lindner}}\ and\ \bibinfo {author} {\bibfnamefont {T.}~\bibnamefont
  {Rudolph}},\ }\href {https://doi.org/10.1103/PhysRevLett.103.113602}
  {\bibfield  {journal} {\bibinfo  {journal} {Phys. Rev. Lett.}\ }\textbf
  {\bibinfo {volume} {103}},\ \bibinfo {pages} {113602} (\bibinfo {year}
  {2009})}\BibitemShut {NoStop}%
\bibitem [{\citenamefont {Economou}\ \emph {et~al.}(2010)\citenamefont
  {Economou}, \citenamefont {Lindner},\ and\ \citenamefont
  {Rudolph}}]{economou2010optically}%
  \BibitemOpen
  \bibfield  {author} {\bibinfo {author} {\bibfnamefont {S.~E.}\ \bibnamefont
  {Economou}}, \bibinfo {author} {\bibfnamefont {N.}~\bibnamefont {Lindner}},\
  and\ \bibinfo {author} {\bibfnamefont {T.}~\bibnamefont {Rudolph}},\ }\href
  {https://journals.aps.org/prl/abstract/10.1103/PhysRevLett.105.093601}
  {\bibfield  {journal} {\bibinfo  {journal} {Phys. Rev. Lett.}\ }\textbf
  {\bibinfo {volume} {105}},\ \bibinfo {pages} {093601} (\bibinfo {year}
  {2010})}\BibitemShut {NoStop}%
\bibitem [{\citenamefont {Gimeno-Segovia}\ \emph {et~al.}(2019)\citenamefont
  {Gimeno-Segovia}, \citenamefont {Rudolph},\ and\ \citenamefont
  {Economou}}]{gimeno2019deterministic}%
  \BibitemOpen
  \bibfield  {author} {\bibinfo {author} {\bibfnamefont {M.}~\bibnamefont
  {Gimeno-Segovia}}, \bibinfo {author} {\bibfnamefont {T.}~\bibnamefont
  {Rudolph}},\ and\ \bibinfo {author} {\bibfnamefont {S.~E.}\ \bibnamefont
  {Economou}},\ }\href
  {https://journals.aps.org/prl/abstract/10.1103/PhysRevLett.123.070501}
  {\bibfield  {journal} {\bibinfo  {journal} {Phys. Rev. Lett.}\ }\textbf
  {\bibinfo {volume} {123}},\ \bibinfo {pages} {070501} (\bibinfo {year}
  {2019})}\BibitemShut {NoStop}%
\bibitem [{\citenamefont {Schwartz}\ \emph {et~al.}(2016)\citenamefont
  {Schwartz}, \citenamefont {Cogan}, \citenamefont {Schmidgall}, \citenamefont
  {Don}, \citenamefont {Gantz}, \citenamefont {Kenneth}, \citenamefont
  {Lindner},\ and\ \citenamefont {Gershoni}}]{schwartz2016deterministic}%
  \BibitemOpen
  \bibfield  {author} {\bibinfo {author} {\bibfnamefont {I.}~\bibnamefont
  {Schwartz}}, \bibinfo {author} {\bibfnamefont {D.}~\bibnamefont {Cogan}},
  \bibinfo {author} {\bibfnamefont {E.~R.}\ \bibnamefont {Schmidgall}},
  \bibinfo {author} {\bibfnamefont {Y.}~\bibnamefont {Don}}, \bibinfo {author}
  {\bibfnamefont {L.}~\bibnamefont {Gantz}}, \bibinfo {author} {\bibfnamefont
  {O.}~\bibnamefont {Kenneth}}, \bibinfo {author} {\bibfnamefont {N.~H.}\
  \bibnamefont {Lindner}},\ and\ \bibinfo {author} {\bibfnamefont
  {D.}~\bibnamefont {Gershoni}},\ }\href
  {https://www.science.org/doi/abs/10.1126/science.aah4758} {\bibfield
  {journal} {\bibinfo  {journal} {Science}\ }\textbf {\bibinfo {volume}
  {354}},\ \bibinfo {pages} {434} (\bibinfo {year} {2016})}\BibitemShut
  {NoStop}%
\bibitem [{\citenamefont {Lee}\ \emph {et~al.}(2019)\citenamefont {Lee},
  \citenamefont {Villa}, \citenamefont {Bennett}, \citenamefont {Stevenson},
  \citenamefont {Ellis}, \citenamefont {Farrer}, \citenamefont {Ritchie},\ and\
  \citenamefont {Shields}}]{lee2019quantum}%
  \BibitemOpen
  \bibfield  {author} {\bibinfo {author} {\bibfnamefont {J.}~\bibnamefont
  {Lee}}, \bibinfo {author} {\bibfnamefont {B.}~\bibnamefont {Villa}}, \bibinfo
  {author} {\bibfnamefont {A.}~\bibnamefont {Bennett}}, \bibinfo {author}
  {\bibfnamefont {R.}~\bibnamefont {Stevenson}}, \bibinfo {author}
  {\bibfnamefont {D.}~\bibnamefont {Ellis}}, \bibinfo {author} {\bibfnamefont
  {I.}~\bibnamefont {Farrer}}, \bibinfo {author} {\bibfnamefont
  {D.}~\bibnamefont {Ritchie}},\ and\ \bibinfo {author} {\bibfnamefont
  {A.}~\bibnamefont {Shields}},\ }\href
  {https://iopscience.iop.org/article/10.1088/2058-9565/ab0a9b/meta} {\bibfield
   {journal} {\bibinfo  {journal} {Quantum Science and Technology}\ }\textbf
  {\bibinfo {volume} {4}},\ \bibinfo {pages} {025011} (\bibinfo {year}
  {2019})}\BibitemShut {NoStop}%
\bibitem [{\citenamefont {Thomas}\ \emph {et~al.}(2022)\citenamefont {Thomas},
  \citenamefont {Ruscio}, \citenamefont {Morin},\ and\ \citenamefont
  {Rempe}}]{thomas2022efficient}%
  \BibitemOpen
  \bibfield  {author} {\bibinfo {author} {\bibfnamefont {P.}~\bibnamefont
  {Thomas}}, \bibinfo {author} {\bibfnamefont {L.}~\bibnamefont {Ruscio}},
  \bibinfo {author} {\bibfnamefont {O.}~\bibnamefont {Morin}},\ and\ \bibinfo
  {author} {\bibfnamefont {G.}~\bibnamefont {Rempe}},\ }\href
  {https://www.nature.com/articles/s41586-022-04987-5} {\bibfield  {journal}
  {\bibinfo  {journal} {Nature}\ }\textbf {\bibinfo {volume} {608}},\ \bibinfo
  {pages} {677} (\bibinfo {year} {2022})}\BibitemShut {NoStop}%
\bibitem [{\citenamefont {Cogan}\ \emph {et~al.}(2023)\citenamefont {Cogan},
  \citenamefont {Su}, \citenamefont {Kenneth},\ and\ \citenamefont
  {Gershoni}}]{cogan2023deterministic}%
  \BibitemOpen
  \bibfield  {author} {\bibinfo {author} {\bibfnamefont {D.}~\bibnamefont
  {Cogan}}, \bibinfo {author} {\bibfnamefont {Z.-E.}\ \bibnamefont {Su}},
  \bibinfo {author} {\bibfnamefont {O.}~\bibnamefont {Kenneth}},\ and\ \bibinfo
  {author} {\bibfnamefont {D.}~\bibnamefont {Gershoni}},\ }\href
  {https://www.nature.com/articles/s41566-022-01152-2} {\bibfield  {journal}
  {\bibinfo  {journal} {Nature Photonics}\ }\textbf {\bibinfo {volume} {17}},\
  \bibinfo {pages} {324} (\bibinfo {year} {2023})}\BibitemShut {NoStop}%
\bibitem [{\citenamefont {Varnava}\ \emph {et~al.}(2008)\citenamefont
  {Varnava}, \citenamefont {Browne},\ and\ \citenamefont
  {Rudolph}}]{varnava2008good}%
  \BibitemOpen
  \bibfield  {author} {\bibinfo {author} {\bibfnamefont {M.}~\bibnamefont
  {Varnava}}, \bibinfo {author} {\bibfnamefont {D.~E.}\ \bibnamefont
  {Browne}},\ and\ \bibinfo {author} {\bibfnamefont {T.}~\bibnamefont
  {Rudolph}},\ }\href {https://doi.org/10.1103/PhysRevLett.100.060502}
  {\bibfield  {journal} {\bibinfo  {journal} {Phys. Rev. Lett.}\ }\textbf
  {\bibinfo {volume} {100}},\ \bibinfo {pages} {060502} (\bibinfo {year}
  {2008})}\BibitemShut {NoStop}%
\bibitem [{\citenamefont {Gubarev}\ \emph {et~al.}(2020)\citenamefont
  {Gubarev}, \citenamefont {Dyakonov}, \citenamefont {Saygin}, \citenamefont
  {Struchalin}, \citenamefont {Straupe},\ and\ \citenamefont
  {Kulik}}]{gubarev2020improved}%
  \BibitemOpen
  \bibfield  {author} {\bibinfo {author} {\bibfnamefont {F.~V.}\ \bibnamefont
  {Gubarev}}, \bibinfo {author} {\bibfnamefont {I.~V.}\ \bibnamefont
  {Dyakonov}}, \bibinfo {author} {\bibfnamefont {M.~Y.}\ \bibnamefont
  {Saygin}}, \bibinfo {author} {\bibfnamefont {G.~I.}\ \bibnamefont
  {Struchalin}}, \bibinfo {author} {\bibfnamefont {S.~S.}\ \bibnamefont
  {Straupe}},\ and\ \bibinfo {author} {\bibfnamefont {S.~P.}\ \bibnamefont
  {Kulik}},\ }\href {https://doi.org/10.1103/PhysRevA.102.012604} {\bibfield
  {journal} {\bibinfo  {journal} {Phys. Rev. A}\ }\textbf {\bibinfo {volume}
  {102}},\ \bibinfo {pages} {012604} (\bibinfo {year} {2020})}\BibitemShut
  {NoStop}%
\bibitem [{\citenamefont {Senellart}\ \emph {et~al.}(2017)\citenamefont
  {Senellart}, \citenamefont {Solomon},\ and\ \citenamefont
  {White}}]{senellart2017high}%
  \BibitemOpen
  \bibfield  {author} {\bibinfo {author} {\bibfnamefont {P.}~\bibnamefont
  {Senellart}}, \bibinfo {author} {\bibfnamefont {G.}~\bibnamefont {Solomon}},\
  and\ \bibinfo {author} {\bibfnamefont {A.}~\bibnamefont {White}},\ }\href
  {https://www.nature.com/articles/nnano.2017.218} {\bibfield  {journal}
  {\bibinfo  {journal} {Nature Nanotechnology}\ }\textbf {\bibinfo {volume}
  {12}},\ \bibinfo {pages} {1026} (\bibinfo {year} {2017})}\BibitemShut
  {NoStop}%
\bibitem [{\citenamefont {Weber}\ \emph {et~al.}(2019)\citenamefont {Weber},
  \citenamefont {Kambs}, \citenamefont {Kettler}, \citenamefont {Kern},
  \citenamefont {Maisch}, \citenamefont {Vural}, \citenamefont {Jetter},
  \citenamefont {Portalupi}, \citenamefont {Becher},\ and\ \citenamefont
  {Michler}}]{weber2019two}%
  \BibitemOpen
  \bibfield  {author} {\bibinfo {author} {\bibfnamefont {J.~H.}\ \bibnamefont
  {Weber}}, \bibinfo {author} {\bibfnamefont {B.}~\bibnamefont {Kambs}},
  \bibinfo {author} {\bibfnamefont {J.}~\bibnamefont {Kettler}}, \bibinfo
  {author} {\bibfnamefont {S.}~\bibnamefont {Kern}}, \bibinfo {author}
  {\bibfnamefont {J.}~\bibnamefont {Maisch}}, \bibinfo {author} {\bibfnamefont
  {H.}~\bibnamefont {Vural}}, \bibinfo {author} {\bibfnamefont
  {M.}~\bibnamefont {Jetter}}, \bibinfo {author} {\bibfnamefont {S.~L.}\
  \bibnamefont {Portalupi}}, \bibinfo {author} {\bibfnamefont {C.}~\bibnamefont
  {Becher}},\ and\ \bibinfo {author} {\bibfnamefont {P.}~\bibnamefont
  {Michler}},\ }\href {https://www.nature.com/articles/s41565-018-0279-8}
  {\bibfield  {journal} {\bibinfo  {journal} {Nature Nanotechnology}\ }\textbf
  {\bibinfo {volume} {14}},\ \bibinfo {pages} {23} (\bibinfo {year}
  {2019})}\BibitemShut {NoStop}%
\bibitem [{\citenamefont {You}\ \emph {et~al.}(2022)\citenamefont {You},
  \citenamefont {Zheng}, \citenamefont {Chen}, \citenamefont {Liu},
  \citenamefont {Qin}, \citenamefont {Xu}, \citenamefont {Ge}, \citenamefont
  {Chung}, \citenamefont {Qiao}, \citenamefont {Jiang} \emph
  {et~al.}}]{you2022quantum}%
  \BibitemOpen
  \bibfield  {author} {\bibinfo {author} {\bibfnamefont {X.}~\bibnamefont
  {You}}, \bibinfo {author} {\bibfnamefont {M.-Y.}\ \bibnamefont {Zheng}},
  \bibinfo {author} {\bibfnamefont {S.}~\bibnamefont {Chen}}, \bibinfo {author}
  {\bibfnamefont {R.-Z.}\ \bibnamefont {Liu}}, \bibinfo {author} {\bibfnamefont
  {J.}~\bibnamefont {Qin}}, \bibinfo {author} {\bibfnamefont {M.-C.}\
  \bibnamefont {Xu}}, \bibinfo {author} {\bibfnamefont {Z.-X.}\ \bibnamefont
  {Ge}}, \bibinfo {author} {\bibfnamefont {T.-H.}\ \bibnamefont {Chung}},
  \bibinfo {author} {\bibfnamefont {Y.-K.}\ \bibnamefont {Qiao}}, \bibinfo
  {author} {\bibfnamefont {Y.-F.}\ \bibnamefont {Jiang}}, \emph {et~al.},\
  }\href
  {https://www.spiedigitallibrary.org/journals/advanced-photonics/volume-4/issue-6/066003/Quantum-interference-with-independent-single-photon-sources-over-300km-fiber/10.1117/1.AP.4.6.066003.short}
  {\bibfield  {journal} {\bibinfo  {journal} {Advanced Photonics}\ }\textbf
  {\bibinfo {volume} {4}},\ \bibinfo {pages} {066003} (\bibinfo {year}
  {2022})}\BibitemShut {NoStop}%
\bibitem [{\citenamefont {Wang}\ \emph {et~al.}(2017)\citenamefont {Wang},
  \citenamefont {He}, \citenamefont {Li}, \citenamefont {Su}, \citenamefont
  {Li}, \citenamefont {Huang}, \citenamefont {Ding}, \citenamefont {Chen},
  \citenamefont {Liu}, \citenamefont {Qin} \emph {et~al.}}]{wang2017high}%
  \BibitemOpen
  \bibfield  {author} {\bibinfo {author} {\bibfnamefont {H.}~\bibnamefont
  {Wang}}, \bibinfo {author} {\bibfnamefont {Y.}~\bibnamefont {He}}, \bibinfo
  {author} {\bibfnamefont {Y.-H.}\ \bibnamefont {Li}}, \bibinfo {author}
  {\bibfnamefont {Z.-E.}\ \bibnamefont {Su}}, \bibinfo {author} {\bibfnamefont
  {B.}~\bibnamefont {Li}}, \bibinfo {author} {\bibfnamefont {H.-L.}\
  \bibnamefont {Huang}}, \bibinfo {author} {\bibfnamefont {X.}~\bibnamefont
  {Ding}}, \bibinfo {author} {\bibfnamefont {M.-C.}\ \bibnamefont {Chen}},
  \bibinfo {author} {\bibfnamefont {C.}~\bibnamefont {Liu}}, \bibinfo {author}
  {\bibfnamefont {J.}~\bibnamefont {Qin}}, \emph {et~al.},\ }\href
  {https://www.nature.com/articles/nphoton.2017.63} {\bibfield  {journal}
  {\bibinfo  {journal} {Nature Photonics}\ }\textbf {\bibinfo {volume} {11}},\
  \bibinfo {pages} {361} (\bibinfo {year} {2017})}\BibitemShut {NoStop}%
\bibitem [{\citenamefont {Wang}\ \emph
  {et~al.}(2019{\natexlab{a}})\citenamefont {Wang}, \citenamefont {Qin},
  \citenamefont {Ding}, \citenamefont {Chen}, \citenamefont {Chen},
  \citenamefont {You}, \citenamefont {He}, \citenamefont {Jiang}, \citenamefont
  {You}, \citenamefont {Wang} \emph {et~al.}}]{wang2019boson}%
  \BibitemOpen
  \bibfield  {author} {\bibinfo {author} {\bibfnamefont {H.}~\bibnamefont
  {Wang}}, \bibinfo {author} {\bibfnamefont {J.}~\bibnamefont {Qin}}, \bibinfo
  {author} {\bibfnamefont {X.}~\bibnamefont {Ding}}, \bibinfo {author}
  {\bibfnamefont {M.-C.}\ \bibnamefont {Chen}}, \bibinfo {author}
  {\bibfnamefont {S.}~\bibnamefont {Chen}}, \bibinfo {author} {\bibfnamefont
  {X.}~\bibnamefont {You}}, \bibinfo {author} {\bibfnamefont {Y.-M.}\
  \bibnamefont {He}}, \bibinfo {author} {\bibfnamefont {X.}~\bibnamefont
  {Jiang}}, \bibinfo {author} {\bibfnamefont {L.}~\bibnamefont {You}}, \bibinfo
  {author} {\bibfnamefont {Z.}~\bibnamefont {Wang}}, \emph {et~al.},\ }\href
  {https://journals.aps.org/prl/abstract/10.1103/PhysRevLett.123.250503}
  {\bibfield  {journal} {\bibinfo  {journal} {Phys. Rev. Lett.}\ }\textbf
  {\bibinfo {volume} {123}},\ \bibinfo {pages} {250503} (\bibinfo {year}
  {2019}{\natexlab{a}})}\BibitemShut {NoStop}%
\bibitem [{\citenamefont {Taballione}\ \emph {et~al.}(2021)\citenamefont
  {Taballione}, \citenamefont {van~der Meer}, \citenamefont {Snijders},
  \citenamefont {Hooijschuur}, \citenamefont {Epping}, \citenamefont
  {de~Goede}, \citenamefont {Kassenberg}, \citenamefont {Venderbosch},
  \citenamefont {Toebes}, \citenamefont {van~den Vlekkert} \emph
  {et~al.}}]{taballione2021universal}%
  \BibitemOpen
  \bibfield  {author} {\bibinfo {author} {\bibfnamefont {C.}~\bibnamefont
  {Taballione}}, \bibinfo {author} {\bibfnamefont {R.}~\bibnamefont {van~der
  Meer}}, \bibinfo {author} {\bibfnamefont {H.~J.}\ \bibnamefont {Snijders}},
  \bibinfo {author} {\bibfnamefont {P.}~\bibnamefont {Hooijschuur}}, \bibinfo
  {author} {\bibfnamefont {J.~P.}\ \bibnamefont {Epping}}, \bibinfo {author}
  {\bibfnamefont {M.}~\bibnamefont {de~Goede}}, \bibinfo {author}
  {\bibfnamefont {B.}~\bibnamefont {Kassenberg}}, \bibinfo {author}
  {\bibfnamefont {P.}~\bibnamefont {Venderbosch}}, \bibinfo {author}
  {\bibfnamefont {C.}~\bibnamefont {Toebes}}, \bibinfo {author} {\bibfnamefont
  {H.}~\bibnamefont {van~den Vlekkert}}, \emph {et~al.},\ }\href
  {https://iopscience.iop.org/article/10.1088/2633-4356/ac168c/meta} {\bibfield
   {journal} {\bibinfo  {journal} {Materials for Quantum Technology}\ }\textbf
  {\bibinfo {volume} {1}},\ \bibinfo {pages} {035002} (\bibinfo {year}
  {2021})}\BibitemShut {NoStop}%
\bibitem [{\citenamefont {Wang}\ \emph
  {et~al.}(2019{\natexlab{b}})\citenamefont {Wang}, \citenamefont {He},
  \citenamefont {Chung}, \citenamefont {Hu}, \citenamefont {Yu}, \citenamefont
  {Chen}, \citenamefont {Ding}, \citenamefont {Chen}, \citenamefont {Qin},
  \citenamefont {Yang} \emph {et~al.}}]{wang2019towards}%
  \BibitemOpen
  \bibfield  {author} {\bibinfo {author} {\bibfnamefont {H.}~\bibnamefont
  {Wang}}, \bibinfo {author} {\bibfnamefont {Y.-M.}\ \bibnamefont {He}},
  \bibinfo {author} {\bibfnamefont {T.-H.}\ \bibnamefont {Chung}}, \bibinfo
  {author} {\bibfnamefont {H.}~\bibnamefont {Hu}}, \bibinfo {author}
  {\bibfnamefont {Y.}~\bibnamefont {Yu}}, \bibinfo {author} {\bibfnamefont
  {S.}~\bibnamefont {Chen}}, \bibinfo {author} {\bibfnamefont {X.}~\bibnamefont
  {Ding}}, \bibinfo {author} {\bibfnamefont {M.-C.}\ \bibnamefont {Chen}},
  \bibinfo {author} {\bibfnamefont {J.}~\bibnamefont {Qin}}, \bibinfo {author}
  {\bibfnamefont {X.}~\bibnamefont {Yang}}, \emph {et~al.},\ }\href
  {https://www.nature.com/articles/s41566-019-0494-3} {\bibfield  {journal}
  {\bibinfo  {journal} {Nature Photonics}\ }\textbf {\bibinfo {volume} {13}},\
  \bibinfo {pages} {770} (\bibinfo {year} {2019}{\natexlab{b}})}\BibitemShut
  {NoStop}%
\bibitem [{\citenamefont {Tomm}\ \emph {et~al.}(2021)\citenamefont {Tomm},
  \citenamefont {Javadi}, \citenamefont {Antoniadis}, \citenamefont {Najer},
  \citenamefont {L{\"o}bl}, \citenamefont {Korsch}, \citenamefont {Schott},
  \citenamefont {Valentin}, \citenamefont {Wieck}, \citenamefont {Ludwig} \emph
  {et~al.}}]{tomm2021bright}%
  \BibitemOpen
  \bibfield  {author} {\bibinfo {author} {\bibfnamefont {N.}~\bibnamefont
  {Tomm}}, \bibinfo {author} {\bibfnamefont {A.}~\bibnamefont {Javadi}},
  \bibinfo {author} {\bibfnamefont {N.~O.}\ \bibnamefont {Antoniadis}},
  \bibinfo {author} {\bibfnamefont {D.}~\bibnamefont {Najer}}, \bibinfo
  {author} {\bibfnamefont {M.~C.}\ \bibnamefont {L{\"o}bl}}, \bibinfo {author}
  {\bibfnamefont {A.~R.}\ \bibnamefont {Korsch}}, \bibinfo {author}
  {\bibfnamefont {R.}~\bibnamefont {Schott}}, \bibinfo {author} {\bibfnamefont
  {S.~R.}\ \bibnamefont {Valentin}}, \bibinfo {author} {\bibfnamefont {A.~D.}\
  \bibnamefont {Wieck}}, \bibinfo {author} {\bibfnamefont {A.}~\bibnamefont
  {Ludwig}}, \emph {et~al.},\ }\href
  {https://doi.org/10.1038/s41565-020-00831-x} {\bibfield  {journal} {\bibinfo
  {journal} {Nature Nanotechnology}\ }\textbf {\bibinfo {volume} {16}},\
  \bibinfo {pages} {399} (\bibinfo {year} {2021})}\BibitemShut {NoStop}%
\bibitem [{\citenamefont {Hanbury~Brown}\ and\ \citenamefont
  {Twiss}(1956)}]{hanbury1956question}%
  \BibitemOpen
  \bibfield  {author} {\bibinfo {author} {\bibfnamefont {R.}~\bibnamefont
  {Hanbury~Brown}}\ and\ \bibinfo {author} {\bibfnamefont {R.}~\bibnamefont
  {Twiss}},\ }\href
  {https://link.springer.com/content/pdf/10.1038/1781447a0.pdf} {\bibfield
  {journal} {\bibinfo  {journal} {Nature}\ }\textbf {\bibinfo {volume} {178}},\
  \bibinfo {pages} {1447} (\bibinfo {year} {1956})}\BibitemShut {NoStop}%
\bibitem [{\citenamefont {Hong}\ \emph {et~al.}(1987)\citenamefont {Hong},
  \citenamefont {Ou},\ and\ \citenamefont {Mandel}}]{hong1987measurement}%
  \BibitemOpen
  \bibfield  {author} {\bibinfo {author} {\bibfnamefont {C.~K.}\ \bibnamefont
  {Hong}}, \bibinfo {author} {\bibfnamefont {Z.~Y.}\ \bibnamefont {Ou}},\ and\
  \bibinfo {author} {\bibfnamefont {L.}~\bibnamefont {Mandel}},\ }\href
  {https://doi.org/10.1103/PhysRevLett.59.2044} {\bibfield  {journal} {\bibinfo
   {journal} {Phys. Rev. Lett.}\ }\textbf {\bibinfo {volume} {59}},\ \bibinfo
  {pages} {2044} (\bibinfo {year} {1987})}\BibitemShut {NoStop}%
\bibitem [{\citenamefont {G{\"u}hne}\ and\ \citenamefont
  {T{\'o}th}(2009)}]{guhne2009entanglement}%
  \BibitemOpen
  \bibfield  {author} {\bibinfo {author} {\bibfnamefont {O.}~\bibnamefont
  {G{\"u}hne}}\ and\ \bibinfo {author} {\bibfnamefont {G.}~\bibnamefont
  {T{\'o}th}},\ }\href {https://doi.org/10.1016/j.physrep.2009.02.004}
  {\bibfield  {journal} {\bibinfo  {journal} {Physics Reports}\ }\textbf
  {\bibinfo {volume} {474}},\ \bibinfo {pages} {1} (\bibinfo {year}
  {2009})}\BibitemShut {NoStop}%
\bibitem [{\citenamefont {Zhong}\ \emph {et~al.}(2018)\citenamefont {Zhong},
  \citenamefont {Li}, \citenamefont {Li}, \citenamefont {Peng}, \citenamefont
  {Su}, \citenamefont {Hu}, \citenamefont {He}, \citenamefont {Ding},
  \citenamefont {Zhang}, \citenamefont {Li} \emph {et~al.}}]{zhong201812}%
  \BibitemOpen
  \bibfield  {author} {\bibinfo {author} {\bibfnamefont {H.-S.}\ \bibnamefont
  {Zhong}}, \bibinfo {author} {\bibfnamefont {Y.}~\bibnamefont {Li}}, \bibinfo
  {author} {\bibfnamefont {W.}~\bibnamefont {Li}}, \bibinfo {author}
  {\bibfnamefont {L.-C.}\ \bibnamefont {Peng}}, \bibinfo {author}
  {\bibfnamefont {Z.-E.}\ \bibnamefont {Su}}, \bibinfo {author} {\bibfnamefont
  {Y.}~\bibnamefont {Hu}}, \bibinfo {author} {\bibfnamefont {Y.-M.}\
  \bibnamefont {He}}, \bibinfo {author} {\bibfnamefont {X.}~\bibnamefont
  {Ding}}, \bibinfo {author} {\bibfnamefont {W.}~\bibnamefont {Zhang}},
  \bibinfo {author} {\bibfnamefont {H.}~\bibnamefont {Li}}, \emph {et~al.},\
  }\href {https://journals.aps.org/prl/abstract/10.1103/PhysRevLett.121.250505}
  {\bibfield  {journal} {\bibinfo  {journal} {Phys. Rev. Lett.}\ }\textbf
  {\bibinfo {volume} {121}},\ \bibinfo {pages} {250505} (\bibinfo {year}
  {2018})}\BibitemShut {NoStop}%
\end{thebibliography}%

\end{document}